\documentclass{article} 
\usepackage{graphicx}
\usepackage{amssymb}
\begin{document}
\title{Towards a strong coupling theory for the KPZ equation}
\author{Hans C. Fogedby\\
\footnote{permanent address} Institute of Physics and Astronomy\\
University of Aarhus, DK-8000, Aarhus C, Denmark\\
and\\
NORDITA\\
Blegdamsvej 17, DK-2100, Copenhagen {\O}, Denmark
}

\date{}
\maketitle
\begin{abstract}
After a brief introduction we review the nonperturbative 
weak noise approach
to the KPZ equation in one dimension. We argue that the
strong coupling aspects of the KPZ equation are
related to the existence of localized
propagating domain walls or solitons representing
the growth modes; the statistical weight of the
excitations is governed by a dynamical action
representing the generalization of the Boltzmann factor
to kinetics.
This picture is not limited to one dimension.
We thus attempt a generalization to higher dimensions
where the strong coupling aspects presumably are
associated with a cellular network of domain walls.
Based on this picture we present arguments for the Wolf-Kertez
expression 
$z= (2d+1)/(d+1)$ for the
dynamical exponent.


\end{abstract}
\section{Introduction}
\label{int}

Driven systems far from equilibrium constitute an enormous
class of natural phenomena including
turbulence in fluids, interface and growth problems,
chemical reactions, processes in glasses and amorphous systems,
biological processes, and even aspects of economical
and sociological structures.

In recent years much of the focus of modern statistical physics
and soft condensed matter has shifted towards such systems. Drawing on 
the case of static and dynamic critical phenomena in and close to
equilibrium, where scaling and the concept of
universality have successfully served to organize our understanding
and to provide a variety of calculational tools, a similar strategy
has been advanced towards the much larger class of nonequilibrium
phenomena with the purpose of elucidating scaling properties and more
generally the morphology or pattern formation in a driven nonequilibrium
state.

In this context the
Kardar-Parisi-Zhang (KPZ) equation, describing the nonequilibrium growth
of a noise-driven interface, provides a simple continuum model
of an open driven nonlinear system exhibiting scaling and pattern
formation \cite{general}.
The KPZ equation for the time evolution of the height $h(\vec x,t)$
has the form \cite{Kardar}
\begin{eqnarray}
&&\frac{\partial h(\vec x,t)}{\partial t} = \nu\nabla^2h(\vec x,t)
+ \frac{\lambda}{2}\vec\nabla h(\vec x,t)\cdot\vec\nabla h(\vec x,t)
+\eta(\vec x,t)-F~,
\label{kpz}
\\
&&\langle\eta(\vec x,t)\eta(\vec x',t')\rangle =
\Delta\delta(\vec x-\vec x')\delta(t-t')~.
\label{noise}
\end{eqnarray}
Here $\nu$ characterizes the damping term, $\lambda$ controls the growth
term, $F$ is a general drift term, and $\eta(\vec x,t)$  a locally
correlated white Gaussian noise modelling the stochastic nature of
the drive or environment; the noise correlations are
characterized by the noise strength $\Delta$.

Starting from a given initial interface
the noise drives after a transient period the interface into a stationary
fluctuating state. Choosing a co-moving frame by setting
$F=(\lambda/2)\langle\vec\nabla h\cdot\vec\nabla h\rangle$ the mean 
height $\langle h\rangle$ decays to zero but the fluctuation spectrum
about $\langle h\rangle=0$, characterized by the moments 
$\langle h^n\rangle$ and generally by the distribution $P(\{h(x)\},t)$, assumes a nontrivial and largely unknown form.
In terms of the local slope $\vec u=\vec\nabla h$ the KPZ equation
assumes the form of the Burgers equation driven by conserved noise
$
\left(\partial/\partial t-(\vec u\cdot\vec\nabla)\right)\vec u
=\nu\nabla^2\vec u+\vec\nabla\eta
\label{burgers}
$
which within the framework of perturbative dynamic renormalization
group (DRG) theory was first studied  
as a model of
noise-driven irrotational fluid flow \cite{Forster}.
This analysis 
was reviewed and extended considerably 
in the context of the KPZ equation \cite{Kardar}. 
The DRG analysis
basically addresses the long time - large distance scaling properties
as illustrated for example by the dynamic scaling hypothesis
\cite{general}
applied to the stationary height-height correlations
\begin{eqnarray} 
\langle(h(\vec x,t) - h(\vec x',t'))^2\rangle
= |\vec x-\vec x'|^{2\zeta}G(|t-t'|/|\vec x-\vec x'|^z)~.
\label{scaling}
\end{eqnarray}
Here the roughness exponent $\zeta$, the dynamic exponent $z$,
and the scaling function $G$ together define the KPZ universality class.
Power counting or a DRG analysis identifies $d=2$ as the lower
critical dimension. An $\epsilon$-expansion about $d=2$ yields
a kinetic phase transition at a finite growth strength $\lambda$ above
$d=2$ from a weak coupling smooth phase with $z=2$ and $\zeta =0$ to
a strong coupling rough phase
with largely unknown scaling exponents. 

In $d=1$
the DRG analysis fortuitously yields $z=3/2$ and $\zeta= 1/2$.
This result also follows from the Galilean invariance of (\ref{kpz}),
$\vec x\rightarrow\vec x-\lambda\vec u_0t$, 
$\vec u\rightarrow\vec u +\vec u_0$, $h\rightarrow h + \vec u_0\cdot\vec x$,
which since
$\lambda$ is unrenormalized under the DRG, implies the scaling law
$z + \zeta =2$.
Together with the known stationary distribution, $P_{\mbox{st}}=
\exp(-(\nu/\Delta)\int dx~u^2)$ \cite{general}, of independent 
slope fluctuations,
yielding $\zeta = 1/2$, we then obtain $z=3/2$. However, unlike critical 
phenomena, the DRG analysis even carried to orders beyond first
loop order \cite{Frey} fails to 
yield detailed insight into the scaling
and pattern formation mechanism in the KPZ equation.

\begin{figure}
\centering
\includegraphics[height=6cm]{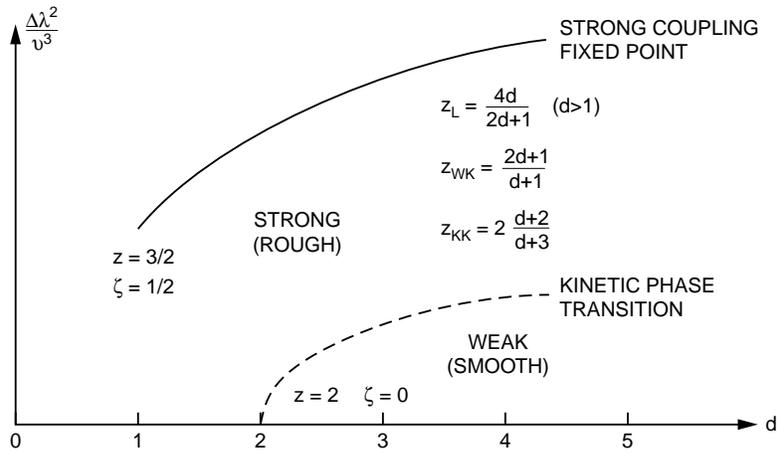}
\caption{
We summarize the scaling properties of the KPZ equation in a plot
of the renormalized coupling constant (the fixed point value)
$\Delta\lambda^2/\nu^3$
versus the dimension $d$ of the system. Below $d=2$ the scaling properties
are determined by the strong coupling fixed. Above $d=2$ the system
exhibits a kinetic phase transition from a weak coupling phase to 
a strong coupling phase. We have also indicated three conjectures
for the exponent $z$: The operator expansion conjecture $z_L$ 
by L\"{a}ssig \cite{Laessig98a} and the conjectures based on numerics,
$z_{WK}$ by Wolf and Kert\'{e}sz \cite{Wolf87} and $z_{KK}$
by Kim and Kosterlitz \cite{Kim89}.
}
\label{fig1}
\end{figure}

A mode coupling analysis (MC) of the KPZ equation has also
been carried out both with regard to a determination of the scaling
function $G$ in $d=1$ and in  search for an elusive upper
critical dimension assumed to be at $d=4$ beyond which 
$z=2$ and $\zeta =0$ \cite{mode}. 
Of other approaches to the scaling properties associated
with the KPZ universality class we mention briefly i) the mapping of
the KPZ equation to the problem of directed polymers in
a quenched random environment \cite{general}, a problem 
in ill-condensed matter,
ii) exact results obtained
for the  $d=1$ asymmetric 
exclusion model \cite{Derrida}  which is 
in the same universality class as the KPZ equation, and finally
iii) recent exact results for a discrete $d=1$ polynuclear growth model
in the KPZ universality class through the association with 
random matrix theory \cite{Praehofer00a}.
In Fig.~\ref{fig1} we have in a plot of the renormalized
coupling strength  versus the dimension summarized the scaling properties
of the KPZ equation.

In summary, the KPZ equation as a continuum
model of an intrinsic nonequilibrium problem and its relationship
to problems in discrete driven lattice gases, random systems of the
spin glass type, and random matrix theory, is of broad and paradigmatic
interest. 
%
\section{Strong coupling theory}
\label{strong}
In recent work we have proposed yet another approach to the
scaling and pattern formation of the KPZ equation
\cite{Fogedby} which we review below.
The strong coupling approach is developed by focusing on the 
Fokker-Planck equation
\begin{eqnarray}
\Delta\frac{\partial P(h(\vec x),t)}{\partial t} =
H P(h(\vec x),t)~,~~H=\int d\vec x~ {\cal H}(\vec x) ~,
\label{fokker}
\end{eqnarray}
associated with the KPZ equation (\ref{kpz}). Here the Hamiltonian
or Liouvillian $H$ drives the transition probability
$P(h(\vec x,t))$
for the height profile $h(\vec x)$ at time $t$.
The Fokker-Planck equation has the form of
a Schr\"{o}dinger equation in imaginary time for the real
positive ``wave function'' $P(h(\vec x),t)$. The noise strength
$\Delta$ plays the role of an effective ``Planck constant''.
Since the KPZ equation is a stochastic field equation the 
associated Fokker-Planck equation is a vastly more complicated
functional-differential
equation. However, applying the 
WKB approximation  valid in the ``semiclassical''
weak noise limit $\Delta\rightarrow 0$ by setting 
($h_1$ in the initial height configuration)
\begin{eqnarray}
P[h_1(\vec x)\rightarrow h(\vec x),t)]
\propto
\exp\left[-\frac{S(h_1(\vec x)\rightarrow h(\vec x),t)}{\Delta}\right]~,
\label{wkb}
\end{eqnarray}
a ``principle of least action'' becomes applicable and the ``weight function''
or action takes the form,
\begin{eqnarray}
S(h_1(\vec x)\rightarrow h(\vec x),t)
=
\int_{h_1(\vec x), t=0}^{h(\vec x), t}
d\vec x dt~\left[p(\vec x,t)\frac{\partial h(\vec x,t)}{\partial t} 
- {\cal H}(\vec x)\right]~,
\label{action}
\end{eqnarray}
This implies the ``classical'' Hamilton equations of motion
\begin{eqnarray}
&&
\frac{\partial h(\vec x,t)}{\partial t} = \nu\nabla^2h(\vec x,t)
+ \frac{\lambda}{2}\vec\nabla h(\vec x,t)\cdot\vec\nabla h(\vec x,t)
-F +p(\vec x,t)~,
\label{field1}
\\
&&
\frac{\partial p(\vec x,t)}{\partial t} = -\nu\nabla^2p(\vec x,t)
+ \lambda\vec\nabla p(\vec x,t)\cdot\vec\nabla h(\vec x,t)
+ \lambda p(\vec x,t)\nabla^2 h(\vec x,t)~,
\label{field2}
\end{eqnarray}
derived from the Hamiltonian density
\begin{eqnarray}
{\cal H} = p(\vec x,t)\left[\nu\nabla^2h(\vec x,t)+
\frac{\lambda}{2}\vec\nabla h(\vec x,t)\cdot\vec\nabla h(\vec x,t)
-F+\frac{1}{2}p(\vec x,t)\right]~.
\label{ham}
\end{eqnarray}
The field equation (\ref{field1}) and (\ref{field2}) are of
the diffusive-advective type and replace the KPZ equation.
The deterministic ``noise field'' $p(\vec x,t)$ corresponds  
to the stochastic noise $\eta(\vec x,t)$ in (\ref{kpz}).

The prescription for the analysis of the KPZ equation is
now straightforward. In order to determine
the transition probability from an initial
height profile $h_1(\vec x)$ at time $t=0$ to a final height profile
$h(\vec x)$ at time $t$ we must evaluate the ``classical''
action associated with a ``classical'' orbit from $h_1(\vec x)$
to $h(\vec x)$ traversed in time $t$. The strong coupling
features enters in the determination of the relevant orbits
which should be nonperturbative solutions
of the field equations.
As is well-known from the WKB approximation in quantum
theory and quantum field theory, the weak noise limit has
no direct bearing on the coupling strength.
The action in the present nonequilibrium context plays
the same role as the energy in the Boltzmann factor in
equilibrium.

\begin{figure}
\centering
\includegraphics[height=6cm]{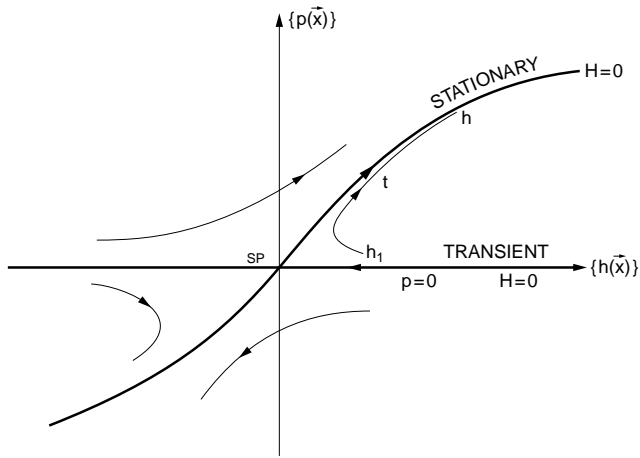}
\caption{
We depict the generic phase space behavior of the WKB
approximation applied to the Fokker-Planck equation.
The stationary saddle point (SP) is at the origin.
The transient zero energy submanifold and
the stationary zero energy submanifold are indicated. At long times
the orbit from $h_1$ to $h$ in time $t$ converges
to the submanifolds and passes close to the saddle point.
}
\label{fig2}
\end{figure}

Since the ``classical'' system is conserved the orbits are
confined to lie on energy surfaces $H=\mbox{const}$. Here
the zero energy surface $H=0$ plays a central role in determining
the stationary distribution $P_{\mbox{st}}(h)=\lim_{t\rightarrow\infty}
P(h_1\rightarrow h,t)$. Unlike the case in mechanical systems, the 
zero energy surface is unbounded and has a characteristic submanifold
structure reflecting the underlying stochastic nature of the
KPZ equation. The transient submanifold for $p=0$ corresponds to
the damped noiseless KPZ equation for $\eta =0$, whereas the 
in general unknown  stationary submanifolds is ``orthogonal'' to
$p=0$ in the sense that the $\int d\vec x{\cal H}=0$ with 
${\cal H}$ given by (\ref{ham}). The submanifolds intersect
in the saddle point $(h,p)=(0,0)$ corresponding to ergodic 
behavior. At long times a specific orbit is unbounded; however, 
an orbit from $h_1$ to $h$ in time $t$ is finite. At long times the orbit
migrates from the finite energy surface to the zero energy submanifolds.
After an initial transient period the noise drives the orbit close
to the saddle point. Ergodic behavior is established and the orbit 
moves out along the stationary submanifold. At long times the orbit
converges to the stationary submanifold and the stationary stochastic
state is attained. This generic phase space behavior is 
depicted schematically in Fig.~\ref{fig2}.
%
\section{Morphology and scaling in one dimension}
\label{one}
The strong coupling WKB analysis
becomes relatively simple in one dimension \cite{Fogedby}. The stationary
zero energy submanifold is given by 
$p(x,t)=-2\nu\nabla^2h(x,t)$, yielding the
stationary distribution \cite{general}
$P_{\mbox{st}}(h)\propto
\exp[-(\nu/\Delta)\int dx(dh/dx)^2]$.
Moreover, the field equations (\ref{field1}) and (\ref{field2})
admit simple cusp solutions, in the static case of the form
$h(x)=(2\nu/\lambda)\log|\cosh((\lambda/2\nu)u(x-x_0))|$ centered about
$x_0$ and for large $x$ approaching the constant slope form
$h\rightarrow u|x-x_0|$. In terms of the local slope $dh/dx$ the
cusps correspond to solitons or domain walls of the kink-like
form $dh/dx=u\tanh((\lambda/2\nu)|u|(x-x_0))$. The right hand soliton
for $u>0$ moves on the $p=0$ submanifold and is the well-known 
viscosity-smoothed
shock wave solution of the noiseless Burgers equation
$(\partial/\partial t-\lambda u\nabla)u=\nu\nabla^2u$. The left
hand soliton for $u<0$ moves on the $p=-2\nu\nabla^2h$ submanifold and is a
solution of the growing Burgers equation 
$(\partial/\partial t-\lambda u\nabla)u=-\nu\nabla^2u$ with negative
damping constant. The cusps or solitons form the elementary
excitations or growth modes in a heuristic many body description of a growing 
interface. Boosting the cusps or solitons to a finite propagation
velocity $v$ by means of the Galilean transformation above 
we obtain the kinematic condition $v=-(2/\lambda)(u_++u_-)$, where
$u_+$ and $u_-$ are the right and left boundary values of the solitons,
respectively. Matching a set of right and left hand solitons according
to the kinematic condition we thus obtain a many body or multi-soliton
representation of a growing interface. 

\begin{figure}
\centering
\includegraphics[height=6cm]{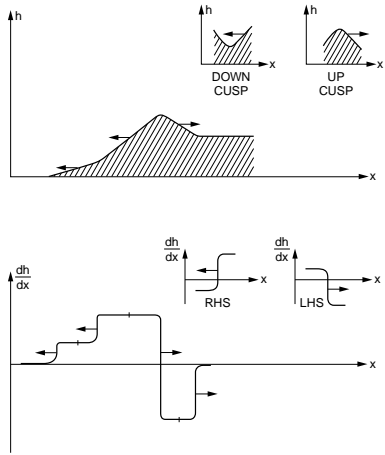}
\caption{
We depict the single cusp and soliton modes
and a multi soliton representation of a growing interface.
We show both the height profile and the slope profile.
}
\label{fig3}
\end{figure}

Imposing periodic boundary conditions it is easily seen by inspection
that the propagation of solitons lead to a growing interface.
A particular simple growth mode corresponds to the propagation of a
pair of matched right and left hand solitons. This mode is equivalent
to the addition of layer upon layer in the evolution of the height
profile.
In Fig.~\ref{fig3} we have shown the single cusps and solitons 
and a multi soliton representation
of a growing interface

The WKB approach also allows us to associate a dynamics
with the propagation of growth modes. Only the noise-induced 
left hand solitons are endowed with dynamical properties.
In terms of the boundary values $u_+$ and $u_-$ we obtain
the Galilean invariant action  $S=(1/6)\lambda\nu|u_+-u_-|^3t$, 
from the Hamiltonian
$H$ the energy 
$E=(2/3)\lambda\nu(u_+^3-u_-^3)$, and from $\Pi=\int dxh\nabla p$
the momentum $\Pi=\nu(u_+^2-u_-^2)$. Considering specifically
a two-soliton mode composed of a right hand and left hand soliton
with amplitude $u$ and with vanishing boundary conditions
it carries action $S=(1/6)|u|^3t$, energy 
$E=-(2/3)\lambda\nu|u|^3$, momentum $\Pi=\nu|u|u$ and propagates
with velocity $v=-\lambda u/2$. Eliminating the velocity we
obtain the soliton dispersion law and from (\ref{wkb}) the
pair soliton distribution, $x$ denotes the center of mass coordinate,
\begin{eqnarray}
E=-\frac{4}{3}\frac{\lambda}{\nu^{1/2}}|\Pi|^{3/2}
~~~\mbox{and}~~~
P(x,t)\propto\exp\left(-\frac{4}{3}\frac{\nu}{\Delta\lambda^2}
\frac{x^3}{t^2}\right)~.
\end{eqnarray}
From these two expressions we draw two results:
i) It follows easily from the spectral properties of the 
height correlations that the exponent $3/2$ in the soliton pair
dispersion law can be identified with the dynamic exponents $z$.
The dynamical scaling is thus associated with the low lying
gapless soliton modes.
ii) In the stochastic growth mode characterized by a gas
of independent pair solitons the solitons perform random
walk with the characteristics of superdiffusion, e.g., 
the distribution implies the mean square displacement
$\langle x^2\rangle(t)\propto(\Delta\lambda^2/\nu)^{1/z}t^{2/z}$
with $z=3/2$.
\section{Towards higher dimensions}
\label{higher}
In dimensions beyond $d=1$ the scaling
and morphology are more involved.
From the DRG and the mapping to
directed polymers if follows that $d=2$ is a lower critical dimension.
Above $d=2$ the system undergoes a kinetic
phase transition at a finite coupling from a weak coupling smooth
phase with exponents $z=2$ and $\zeta=0$ to a strong coupling rough
phase. An operator expansion method
yields $z=4d/(2d+1)$ for $d>1$ 
\cite{Laessig98a}. 
Two other heuristics proposals
are $z=2(d+2)/(d+3)$ \cite{Kim89} and $z=(2d+1)/(d+1)$ \cite{Wolf87} 
based on numerics;
they both yield $z=3/2$ in $d=1$. Common to all these expressions
is that $z\rightarrow 2$ for $d\rightarrow\infty$, i.e., yielding
an infinite upper critical dimension. 
\cite{mode}.

Within the present WKB approach the evolution of the morphology
of the growing interface is given by the appropriate strong coupling
solutions of the ``classical'' equations (\ref{field1}) and 
(\ref{field2}). In one dimension we were able analytically to find soliton
solutions and thus ``parameterize'' a growing interface in terms
of a gas of kinematically matched solitons. We have also analyzed the
equations numerically in one dimension but here the negative diffusion
coefficient in the equation for the noise field $p$ renders the
coupled equations numerically unstable and we were only able to
verify certain soliton collision configurations. The same instability
problem will remain in higher dimension and it is at the moment not
clear how to find solutions numerically. 

Starting from random initial conditions the time evolution of the
noiseless Burgers equation is damped \cite{Woyczynski98}. Since the nonlinear 
Cole-Hopf transformation \cite{general}
relates the slope field to a linear diffusive field the transient morphology
can be analyzed by saddle point methods and is characterized by a 
gas of localized right hand solitons connected by ramp
solutions; for the height field
this pattern formation corresponds to a gas of downward cusps connected
by parabolic segments. In two and higher dimensions one finds a corresponding
morphology \cite{Woyczynski98}. The height field forms a damped 
cellular network of 
growing ``domes''
connected by ``valley's''. The domes and valleys correspond to the 
parabolic segments and downward cusps in one dimension, respectively.

\begin{figure}
\centering
\includegraphics[height=6cm]{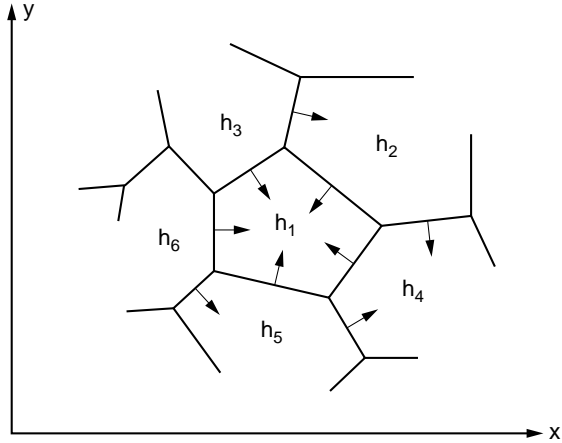}
\caption{
We depict the cellular network of growing and decaying islands
in the height profile. The network is formed of domain walls or solitons
constituting the fundamental growth modes.
}
\label{fig4}
\end{figure}

When the noise is ``turned on'' the right hand soliton in one dimension
is supplemented by a noise induced left hand soliton in the WKB
interpretation and as discussed above the morphology can be
discussed in terms of connected right hand and left hand solitons. For the
height field one version of this morphology corresponds to 
growing and decaying plateaus. The moving steps connecting the plateaus
correspond to soliton pairs in the slope field. Similarly,
we believe that a corresponding morphology takes place in
higher dimensions. In for example two dimensions the stationary
growth morphology in the height field is composed of
growing and decaying islands forming a cellular network. 
The islands are connected by
ramps or steps propagating with a velocity dictated by the
Galilean invariance of the corresponding Burgers equation.
The statistical weight of a morphology is then given by the
action associated with a given time evolution from $h_1$ to
$h$ in a time $t$. If we parameterize the islands by linear
segments and ignore vertex corrections we can make use of the
one dimensional soliton solutions and connect the island
by means of pair-solitons of amplitude $u$. The steps
will then propagate with a velocity $v$ or order $\lambda u$
and carry an action of order $\lambda\nu u^3t$ per. unit length.
In Fig.~\ref{fig4} we have depicted a growth configuration
for the height field in two dimensions.

We now have enough elements of a future more detailed  theory
to venture a proposal for the dynamic exponent $z$.
First we notice that the localized growth modes lie on the
stationary manifold $H=0$. In one dimension this manifold
is given by $p=-2\nu\nabla^2 h$, yielding the action
$S\propto\lambda\nu|u|^3t$ and the stationary distribution
$P_{\mbox{st}}(h)\propto\exp[-(\nu/\Delta)\int dx(dh/dx)^2]$.
In higher dimensions the stationary
manifold is unknown and the stationary distribution is expected to be 
non-Gaussian and skew \cite{Derrida}. However, assuming that 
the Galilee 
invariant action for a step in $h$ is a function of the slope change
$\delta u$ (ignoring the vector character of $\vec u$) and 
expressing $S$ in the form
$S\propto\lambda\nu|u|^\alpha\ell^{d-1}t$, where the factor
$\ell^{d-1}$ arises from the extended character of the higher
dimensional ``solitons'' or ``domain walls'' providing the growth
modes, a natural choice interpolating between $d=1$ and
$d=\infty$ is to set  $\alpha=d+2$.
Finally, noting that $\delta u\propto v$ and balancing the space and 
time dependencies of $v$ and $\ell$ we infer
a dynamic exponent in agreement with 
Wolf and Kertez \cite{Wolf87}
\begin{eqnarray}
z=\frac{2d+1}{d+1}
\end{eqnarray}
This expression agrees with $z=3/2$ in $d=1$. In $d=2$ we obtain
$z=5/3$ which is close to the numerical value \cite{general}. For 
$d\rightarrow\infty$ we have $z\rightarrow 2$ and the analysis
yields an infinite upper critical dimension.

\medskip
\noindent
{\bf\large Acknowledgements}

\noindent
Discussions with J. Krug, M. Moore, A. Bray, M. H. Jensen,
T. Bohr, A. Svane
and K. Rechendorff
are gratefully acknowledged.



\end{document}